# Multi-UE Networked Sensing for 6G Perceptive Mobile Networks

J. Andrew Zhang, Jingying Bao, Kai Wu, Henk Wymeersch, Christos Masouros, and Y. Jay Guo

**Abstract:** Networked sensing, which jointly exploits observations from multiple distributed nodes, is essential for unlocking the full sensing potential of integrated sensing and communications (ISAC). This article presents a systematic framework for multi-UE sensing in future perceptive mobile networks, exploiting the diverse yet correlated sensing observations naturally arising from distributed user equipment devices (UEs) interacting with common targets. We investigate representative uplink, downlink, and hybrid architectures and reveal distinctive sensing opportunities enabled by different multiple-access schemes. A correlation-aware signal processing framework encompassing synchronization, joint sensing parameter estimation, and multi-view fusion is presented, with emphasis on collectively exploiting distributed communication resources. Key research challenges, including correlation modelling, target association, sensing information compression, and communication-sensing co-optimization, are also discussed.

## 1. INTRODUCTION

Future sixth-generation (6G) mobile networks are expected to evolve from communication-centric infrastructures into perceptive mobile networks (PMNs) with integrated sensing and communications (ISAC) capability [1]. ISAC standardization is also advancing in 3GPP, with Release 19 establishing service requirements and sensing channel models, and Release 20 further investigating NR ISAC in TR 38.765. These developments provide a foundation for evaluating emerging sensing architectures and potentially supporting their implementation in future standardised networks.

A key capability required to fully realize PMNs is networked sensing, where sensing observations from multiple distributed wireless nodes are jointly exploited to obtain multi-view, high-resolution perception of the environment. Without cooperation among network nodes, individual sensing devices typically provide only fragmented and limited observations due to restricted spatial perspectives, blockage, and fading. By leveraging distributed observations across the network, networked sensing can substantially improve sensing resolution, robustness, coverage, and target identifiability [2-7].

Existing networked sensing approaches predominantly rely on multiple-cooperative-receiver (multi-CoopRx) architectures, where multiple coordinated receivers, typically base stations (BSs), jointly perform sensing using signals from a single transmitter [2]-[6], as illustrated in Fig. 1(a). While such approaches can improve sensing performance for multiple sensing applications as exemplified in these works, they often require densely deployed BSs and extensive network coordination, including tight inter-BS synchronization, rigorous interference management, and high-capacity backhaul links for exchanging large volumes of sensing data. These requirements significantly increase system complexity, deployment cost, and implementation difficulty, limiting scalability in practical mobile networks. Moreover, the emphasis on BS-side cooperation has left the sensing potential of distributed user equipment devices (UEs) largely underexplored, despite their abundance in modern multi-access wireless networks. Although recent studies have demonstrated the potential of multi-UE sensing in specific applications, such as UE-assisted target localization using joint monostatic and bistatic ranging [7], a systematic framework encompassing general sensing architectures, correlation-aware joint processing, and the collective exploitation of multi-access communication resources remains lacking.

In this article, we develop a general framework for multi-UE networked sensing, with particular emphasis on the commercially attractive multi-UE single-BS multistatic architecture shown in Fig. 1(b). Unlike conventional multi-CoopRx sensing, in this architecture, multiple non-cooperative UEs form distributed bistatic sensing links with a common BS, where sensing information is extracted from uplink signals, downlink signals, or both and jointly exploited for sensing. The framework exploits a fundamental property of wireless networks that has received limited attention: geographically proximate UEs often illuminate or observe the same targets from different spatial perspectives, whether acting as transmitters or receivers, generating multi-view observations with different yet correlated *sensing parameters*, such as delay, Doppler, angles, and amplitudes. By jointly exploiting both multi-view diversity and correlation, signals from multiple UEs can be fused at a single BS to achieve high-resolution

J.A. Zhang (*Senior member, IEEE*), J. Bao, K. Wu, and Y.J. Guo (*Fellow, IEEE*) are with University of Technology Sydney, Australia. Email: {Andrew.Zhang; Kai.Wu; Jay.Guo}@uts.edu.au; Jingying.Bao@student.uts.edu.au. H. Wymeersch (*Fellow, IEEE*) is with Chalmers University of Technology, Sweden. Email: henkw@chalmers.se. C. Masouros (*Fellow, IEEE*) is with University College London, UK. Email: c.masouros@ucl.ac.uk.

This research was partially supported by the Australian Research Council's DP220101158 and by the SNS JU project 6G-DISAC under the EU's Horizon Europe research and innovation program under Grant Agreement No. 101139130. Figs. 1 and 2 are generated with the assistance of ChatGPT.

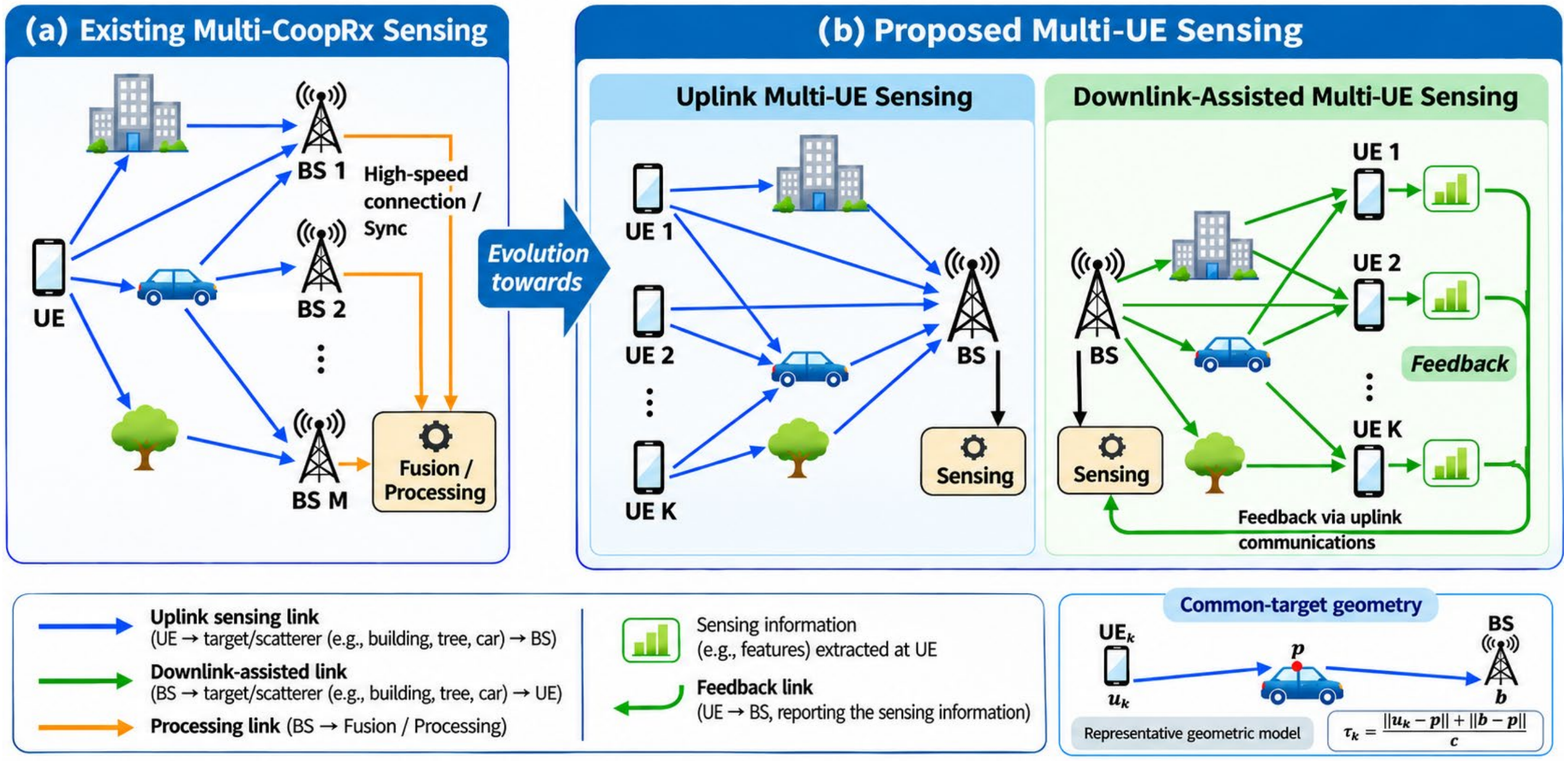

*Fig. 1 Evolution from (a) multi-CoopRx sensing, where sensing diversity and fusion are primarily realized through cooperating BS receivers, to (b) Multi-UE networked sensing, where sensing diversity is generated by distributed UEs without inter-UE cooperation, and fusion can be performed at a single BS. The car is the target being sensed.*

perception with significantly reduced coordination and infrastructure requirements.

The main contributions of this article are threefold. First, we develop a systematic framework for multi-UE networked sensing, encompassing representative uplink, downlink-assisted, and hybrid architectures, and identifying the critical signal processing stages and enabling techniques. Second, we reveal two fundamental and interconnected principles of multi-UE sensing: the exploitation of inter-UE sensing correlation during joint signal processing and the collective exploitation of time, frequency, and spatial resources distributed across UEs through multiple-access communications. Third, we identify key research challenges and opportunities arising from these principles, including correlation modelling, sensing performance characterization, target association, sensing information compression, and correlation-aware UE grouping and communication-sensing co-optimization.

Although this article primarily focuses on the single-BS architecture, the proposed multi-UE sensing framework can be naturally extended to systems with multiple cooperating BSs. In addition to cellular networks, the framework is also applicable to other multi-access wireless systems such as Wi-Fi, low-altitude networks, and LoRa networks.

The remainder of this article is organized as follows. Section 2 presents representative multi-UE sensing architectures. Section 3 discusses key sensing mechanisms and enabling technologies. Section 4 highlights major technical challenges and open research problems. Finally, Section 5 concludes the article.

## 2. Multi-UE Sensing Architectures

As illustrated in Fig. 1(b), the principles of multi-view sensing and correlation exploitation can be realized through several architectural frameworks depending on how sensing observations are acquired and fused across the network. The most representative architectures include uplink sensing, downlink-assisted sensing, and hybrid sensing that jointly exploits both uplink and downlink observations.

### 2.1 Uplink Multi-UE Sensing

In uplink multi-UE sensing, multiple UEs simultaneously transmit communication signals to a BS, while the BS jointly exploits the direct and reflected signal components for sensing. Each UE-BS pair forms a distinct bistatic sensing link, and the spatial distribution of UEs creates multiple sensing geometries for observing common targets. Consequently, the network can obtain multiple views of the same target through transmitter-side spatial diversity. This architecture is particularly attractive because it aligns well with current cellular protocols and naturally supports scalable sensing through centralized BS processing, without requiring UE-side sensing or additional sensing-information reporting. Multi-view observations from geographically distributed UEs can enhance target observability and separability, improve robustness against blockage and fading, and extend sensing coverage. These advantages are particularly valuable in dynamic outdoor environments containing multiple targets, scatterers, and interference sources.

However, uplink sensing typically requires multiple UEs to share communication resources, with each UE occupying only a subset of the available resources or experiencing inter-user interference under spatial multiplexing. While reusing existing uplink transmissions does not inherently introduce additional inter-cell interference, sensing-oriented resource allocation may alter interference to neighbouring BSs. Conversely, inter-cell interference may degrade sensing performance, motivating joint communication-sensing interference management.

| Schemes | Potential Sensing Improvements | Limitations | Coherent Gain Conditions and Mechanisms |
|---|---|---|---|
| TDMA | Improved Doppler resolution and/or enlarged estimation range; Improved tracking | Doppler resolution and estimation range (ambiguity) limited by aggregated channel observation duration and temporal sampling rate, respectively; Limited simultaneous multi-view diversity | Cross-UE Doppler and phase alignment; Temporal alignment and aggregation; Gains constrained by the coherent processing interval. |
| OFDMA | Improved delay resolution and/or enlarged delay estimation range; scalable parallel sensing | Delay resolution and ambiguity limited by aggregated subcarrier span and interval, respectively; Fragmented spectrum resources; scheduling complexity | Cross-UE delay modelling and phase calibration; Geometry-aware delay alignment; Gains constrained by the accuracy of cross-UE delay and phase relationships. |
| SDMA | Improved angular estimation and target separability; potential distributed aperture gains | Angular resolution and ambiguity limited by the spatial extent and spacing of participating UEs, respectively; Limited number of spatially separable UEs; Inter-user interference; High complexity | Geometry-aware phase calibration; Calibrated antenna for angle estimation; Distributed aperture formation; Accurate UE positions; Gains constrained by UE geometry and scattering consistency. |

*Table 1 Impact of multi-access schemes on multi-UE sensing performances. Common benefits such as improved sensing robustness and detection reliability through multi-view diversity are omitted. Coherent gains are potential rather than guaranteed benefits and depend on the underlying propagation geometry, resource allocation, and synchronization and calibration accuracy.*

## 2.2 Downlink-Assisted Multi-UE Sensing

In downlink-assisted multi-UE sensing, distributed UEs receive downlink signals transmitted by one or multiple BSs and locally extract sensing-related features. Rather than forwarding raw received signals, the UEs reduce overhead by reporting compressed measurements or high-level sensing features to the network edge or BS for joint fusion, although additional UE-side processing and feedback are required. This architecture enables the network to exploit sensing observations collected from diverse UE locations and perspectives, substantially enriching the available sensing information.

Compared with uplink-based sensing, downlink sensing can potentially leverage a much larger number of sensing observations due to the broadcast nature of downlink transmissions. A single BS transmission can be simultaneously observed by numerous UEs in a cell, and downlink sensing allows every participating UE to exploit the received signal for sensing, enabling denser spatial sampling and richer multi-view diversity. In addition, the BS location is often known to UEs, important for target localization, whereas the reverse is not necessarily true.

## 2.3 Hybrid Uplink-Downlink Multi-UE Sensing

Hybrid multi-UE sensing simultaneously extracts sensing information from uplink transmissions generated by a set of distributed UEs and from downlink signals observed by the same or a different set of distributed UEs, leveraging the complementary advantages of uplink and downlink-assisted sensing architectures.

Although uplink and downlink sensing observations are governed by the same physical environment, they are not necessarily identical. In frequency-division-duplex (FDD) systems, uplink and downlink operate on different frequency bands and therefore experience different channels. Even in time-division-duplex (TDD) systems, sensing observations may differ due to different resource allocations, beamforming configurations, reference signals, transmission time, and participating nodes. As a result, uplink and downlink sensing often provide complementary measurements, sensing qualities, and observation opportunities. More importantly, hybrid sensing enables simultaneous exploitation of sensing observations collected from both transmitting and receiving UEs, increasing the number of available observations and strengthening the geometric constraints that can be exploited for localization, tracking, and environmental reconstruction.

Uplink and downlink observations can be independently processed and subsequently fused at the parameter or higher levels without requiring cross-link phase coherence. For coherent signal-level combination, necessary additional calibration across links and frequency bands can be performed using the techniques discussed later in Section 3.1.

## 2.4 Multiple-Access-Aware Sensing across Multi-UE Architectures

Previous subsections classify multi-UE sensing by how observations are acquired and processed, this subsection considers the complementary dimension of multiple-access schemes, which determine the sensing resources available to individual UEs, particularly in the uplink. Rather than exploiting these resources independently, multi-UE sensing seeks to allocate them collectively through correlation-aware joint processing. When UE signals, whether pilots or data symbols, are orthogonal or otherwise separable, UE-specific sensing observations can be extracted under different multiple-access schemes, although coherent combination requires further synchronization, calibration, and geometry-aware processing.

As summarized in Table 1, different schemes offer distinct opportunities for collectively exploiting time, frequency, and spatial resources, with different enabling conditions, leading to different sensing benefits and trade-offs. More specifically, time division multiple access (TDMA) provides temporal diversity through additional sensing snapshots, orthogonal frequency division multiple access (OFDMA) enables flexible frequency-domain diversity via subcarrier and bandwidth aggregation, while spatial division multiple access (SDMA) offers spatial and angular diversity through simultaneous observations from different spatial directions.

The signal allocation pattern can further influence sensing performance. Taking OFDMA as an example, different subcarrier allocation strategies may lead to distinct sensing benefits. When each UE occupies localized subcarrier blocks, the aggregated bandwidth across multiple UEs can be exploited to improve delay resolution through stitched wideband sensing. In contrast, interleaved subcarrier allocation increases frequency-domain sampling diversity, which can enlarge the unambiguous delay estimation range and mitigate range ambiguities. Thus, a mixed, non-uniform subcarrier allocation may offer a favourable trade-off between delay resolution and unambiguous delay range. Similar opportunities exist in TDMA and SDMA systems. Consequently, future multi-UE sensing systems may require sensing-aware multiple-access design, where communication resource allocation is jointly optimized with sensing objectives.

## 3. Multi-UE Sensing Signal Processing Framework

Realizing the potential of multi-UE sensing requires a signal processing framework capable of transforming heterogeneous observations collected from distributed UEs into reliable environmental perception. As illustrated in Fig. 2, uplink sensing involves centralized BS processing, including bistatic offset cancellation, correlation-aware parameter estimation, target association, and multi-view fusion. In downlink-assisted sensing, UEs extract and compress sensing information for reporting to the BS, where measurements or features at different levels can be jointly processed and fused, without requiring direct UE-to-UE cooperation. The framework also incorporates UE localization, grouping, and resource optimization, with sensing-driven feedback enabling adaptive processing and resource allocation.

The framework is fundamentally built upon the correlation structure among sensing observations, which originates from the common targets and propagation environment observed by multiple UEs. Such correlation may be represented through channel covariance matrices, joint probability distributions of sensing parameters, and geometric relationships among bistatic sensing paths. Representative asynchronous bistatic sensing models can be found in [8, 9]. To illustrate the underlying correlation, consider the idealized common target geometric model where an uplink system with a BS at position $b$, UE $k$ at position $\boldsymbol{u}_k$, and a common scatterer at position $\boldsymbol{p}$, as illustrated in the bottom-right inset of Fig. 1. The bistatic propagation delay associated with UE $k$ is $(\|\boldsymbol{p}-\boldsymbol{u}_k\|+\|\boldsymbol{p}-\boldsymbol{b}\|)/c$, where c is the speed of light. The spatial separation between two UEs imposes an upper bound on their bistatic propagation delay difference. Similarly, their Doppler shifts are linked through common target motion and UE-dependent sensing directions and velocities, while their observations may share the same angle of arrival (AoA) at the BS. However, UE-dependent clock offsets may obscure these relationships, motivating correlation-preserving offset compensation.

Multi-UE sensing can exploit both non-coherent and coherent sensing diversity. Non-coherent processing exploits diversity and correlation among UE observations to improve sensing coverage, robustness, target separability, and multi-view fusion without requiring stringent inter-UE synchronization. Coherent processing can potentially provide additional gains, including coherent SNR improvement, bandwidth aggregation, and distributed aperture formation, but generally requires accurate synchronization, calibration, and knowledge of UE geometry. The following discussion considers both processing paradigms, with particular emphasis on correlation-aware joint processing and the exploitation of coherent sensing gains under practical synchronization and calibration constraints.

In principle, signals corresponding to both pilots and random data payloads can be exploited for sensing, similar to conventional bistatic sensing [1, 8]. However, the use of data payloads in multi-UE sensing introduces additional challenges. Firstly, it is generally difficult to directly apply conventional radar-like correlation or matched-filter processing to the received time-domain signal, which consists of superimposed transmissions from multiple users propagated through different channels. Consequently, sensing often needs to be performed after transforming the received signal into a domain where different users become separable. Explicit decoding or removal of the data symbols, however, may not always be necessary. Secondly, in SDMA systems, payload-based sensing may be further affected by residual multi-user interference and imperfect user separation. Finally, when the decision-directed method is used, decoding errors and signal reconstruction uncertainties can distort the correlation structure among sensing observations from different UEs.

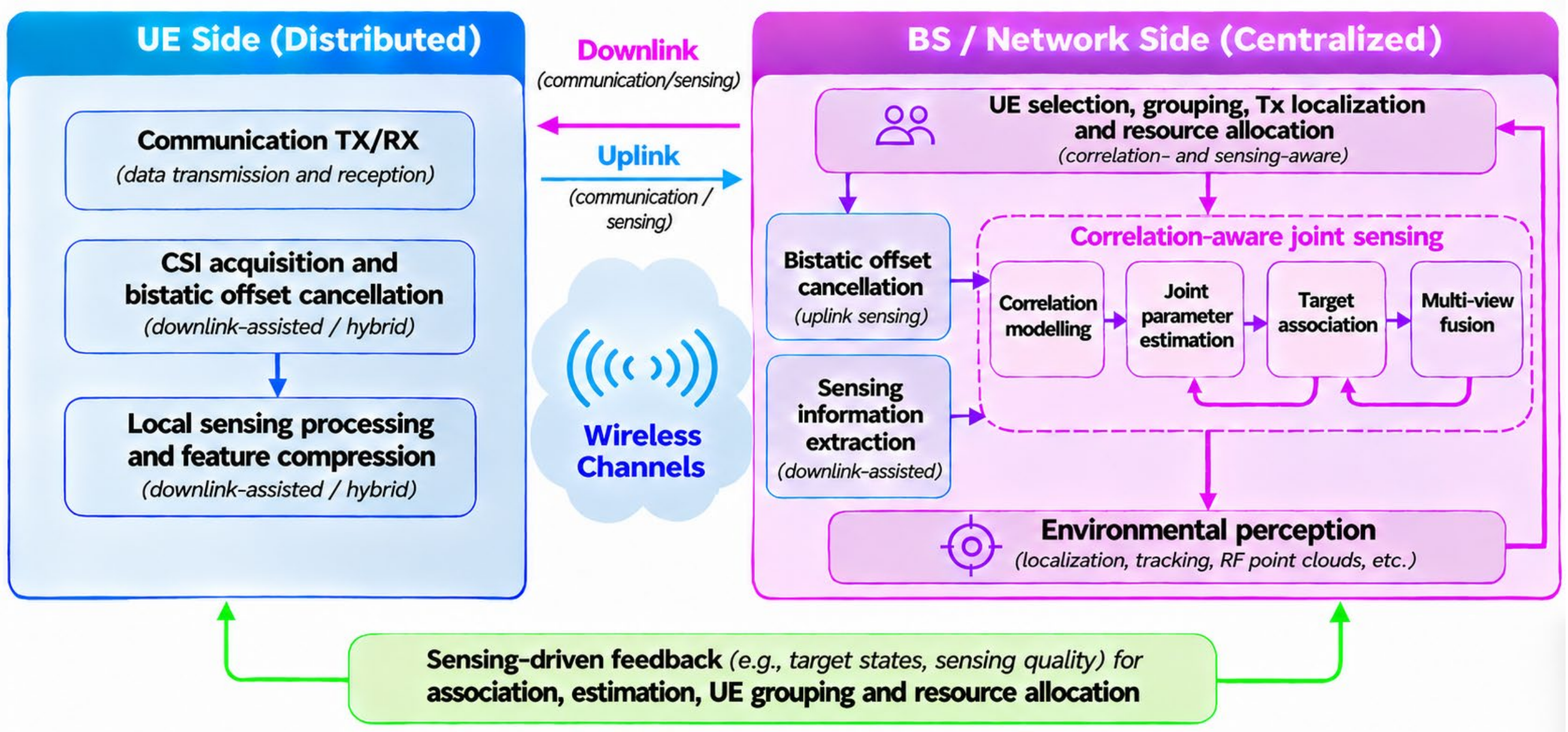

*Fig. 2 Multi-UE networked sensing framework illustrating UE- and BS-side main processing for uplink and downlink-assisted sensing.*

Hereafter, we mainly consider sensing based on channel state information (CSI) estimated from pilots. As illustrated in Fig. 2, three core signal processing functions are essential for exploiting multi-UE observations: bistatic offset cancellation to mitigate clock asynchronism and preserve inter-UE sensing relationships; correlation-aware joint parameter estimation to exploit common-target information across UEs; and multi-view fusion to integrate complementary observations into a comprehensive environmental representation. These functions form the focus of the remainder of this section. Other major interconnected modules are discussed as research challenges in Section 4.

### 3.1 Synchronization and Bistatic Offset Cancellation

Synchronization is one of the fundamental challenges in multi-UE networked sensing. Sensing observations are affected by various offsets induced by clock asynchronism, including carrier frequency offsets, phase offsets, and timing offsets [8]. These offsets are particularly problematic in multi-UE sensing because they vary across UEs and may be significantly larger than the target-induced variations of interest. Without proper compensation, the resulting sensing parameters can be severely distorted, obscuring the geometric relationships among observations from different UEs.

Recent studies have shown that many of these offsets can be effectively mitigated through per-subcarrier processing in multicarrier systems [8]. Representative approaches include cross-correlation, as well as signal-ratio processing, based on CSI between received signals and a locally generated reference signal. The reference signal can be constructed from the received signal, corresponding to a dominant path or signal subspace [9], without requiring a dedicated reference hardware branch. By exploiting their common offsets, these operations can largely eliminate unknown transmitter- and receiver-dependent offsets while preserving relative delay, Doppler, and angular information associated with targets.

The implementation of offset suppression in multi-UE uplink sensing depends on the adopted multiple-access scheme, although the principled approach above can be applied. In TDMA systems, processing can be performed independently for each UE transmission interval. In OFDMA systems, offset cancellation is typically carried out on a per-subcarrier basis before combining observations across frequency resources. In SDMA systems, synchronization processing is often integrated with spatial separation and beamforming operations. Despite these architectural differences, the underlying objective remains the same: transforming heterogeneous multi-UE observations into a common sensing representation that is largely invariant to communication-specific synchronization errors.

However, these operations often convert absolute sensing parameters into relative measurements. Recovering absolute sensing parameters generally requires additional prior information, such as a known reference or anchor path, calibration, or bidirectional signaling. In cellular networks, the direct LoS path between a UE and a BS provides a natural anchor when available, with its delay determined from their positions. Alternatively, stable paths associated with known environmental reflectors may serve as anchors through environmental mapping or prior calibration. Network-assisted positioning can further facilitate anchor identification and calibration. Following offset compensation, weak target-reflected paths can be separated from dominant direct-path and clutter components through joint delay-angle-Doppler processing, reference-path subtraction, or subspace-based filtering [8, 9].

While beneficial for offset suppression, relative sensing parameters may also distort or partially remove the correlation structure required for multi-UE sensing. Therefore, it is crucial to design processing methods that eliminate synchronization-induced offsets while preserving the target-dependent geometric relationships among sensing observations. An example is provided in [10], where the correlation among sensing parameters is retained through the known delay of a constructed reference signal. Extending such approaches to more general scenarios without reliable prior information remains an open research challenge.

While the above techniques address link-specific synchronization offsets, coherent combination of their observations may require further inter-link and inter-band calibration. Multiband sensing techniques [11] provide promising approaches to compensating for inter-band offsets and enabling coherent bandwidth aggregation. Their extension to multi-UE sensing requires additional consideration of UE-dependent sensing geometries and frequency-dependent propagation characteristics. In contrast, parameter- or higher-level fusion does not require cross-link phase coherence.

## 3.2 Sensing Parameter Estimation with Correlation Exploitation

Following synchronization and offset suppression, the next key step is to jointly estimate sensing parameters, such as delay, Doppler, and angles from multi-UE observations. Common targets can be exploited through explicit geometric modelling that links sensing parameters to shared target states [7], or through more general correlation-aware joint estimation that implicitly exploits geometric and statistical relationships among UE signals or CSI [10]. The latter is particularly valuable when individual UEs have limited bandwidth, observation time, or sensing quality due to multi-access resource constraints.

Table 2 compares representative correlation-aware estimation techniques and their applicability conditions. The effectiveness of correlation exploitation depends on sensing geometry and propagation conditions. Geometric approaches require sufficiently accurate UE geometry and tractable common-target relationships, statistical approaches rely on reliable, shared priors, while structured sparsity requires common or partially overlapping supports, potentially after geometry-aware alignment.

The computational complexity of joint estimation, particularly Bayesian methods, may increase substantially with the number of participating UEs and sensing parameters. Exploiting shared geometric and statistical structures can reduce the effective parameter dimension, while approximate Bayesian inference, structured sparse recovery, and sequential processing offer promising approaches to improving computational scalability.

Figure 3 illustrates the benefits of correlation-aware sensing parameter estimation in [10], where the number of jointly processed UEs varies within a 16-UE OFDMA system, with each UE occupying 128 localized subcarriers. By applying

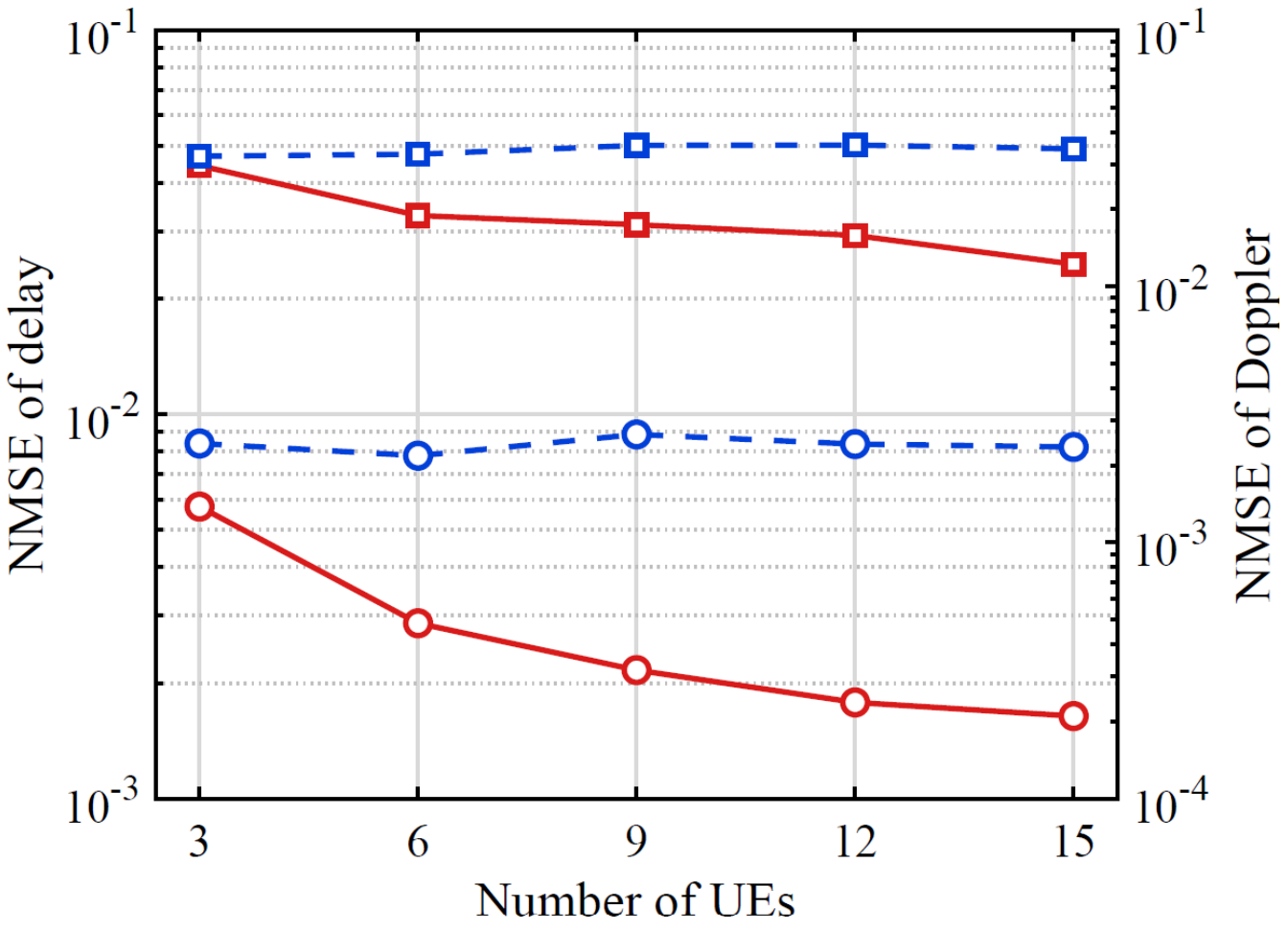

*Fig. 3 NMSE of delay and Doppler estimation versus the number of jointly processed UEs in an uplink OFDMA system with 16 UEs sharing 2048 subcarriers (128 localized subcarriers per UE), SNR = 10 dB, an eight-element BS antenna array, and 60 kHz subcarrier spacing. Solid/dashed curves represent joint/individual-UE processing; square/circle markers denote Doppler/delay estimation. Other simulation settings follow [10].*

Bayesian compressive sensing to jointly exploit the propagation-delay correlation among nearby UEs observing a common target, the BS can achieve substantially higher estimation accuracy than conventional independent per-UE processing approaches. While these results demonstrate the benefits of increasing UE participation under fixed per-UE resource allocations, a fixed total resource budget introduces a trade-off, as the sensing resources available to each UE may decrease with increasing UE participation, potentially offsetting the gains from multi-UE diversity. This may lead to an optimal number of participating UEs and non-monotonic sensing performance. Such trade-offs depend on the multiple-access scheme, sensing geometry, inter-UE channel correlation, and processing algorithms, motivating joint UE grouping and resource optimization.

## 3.3 Multi-view Fusion

Multi-view fusion integrates complementary sensing information from distributed UEs to construct a more comprehensive representation of the environment, including target locations, trajectories, and RF point clouds. While correlation-aware parameter estimation improves the estimation of sensing parameters by exploiting their statistical and geometric relationships, multi-view fusion focuses on combining information from different spatial perspectives to improve target observability, coverage, and environmental reconstruction. For example, complementary viewing geometries may reveal additional scattering points in RF point cloud reconstruction, improving point cloud density and completeness, while correlation-aware parameter estimation improves the accuracy and separability of individual scattering points. These benefits can be characterized by point cloud

| Category | Representative Techniques | Correlation / Prior Exploited | Key Advantages | Key Limitations |
|---|---|---|---|---|
| **Geometry-Constrained Estimation** | Geometric fitting; Joint target-state estimation; Geometry-constrained maximum likelihood (ML) | Common target geometry; Bistatic delay-Doppler-angle relationships across UEs | Exploits common targets despite different UE-specific sensing parameters; Enables joint localization and association | UE geometry uncertainty; Nonlinear and nonconvex estimation; Model mismatch and partial target visibility |
| **Structured Sparse Recovery** | Joint sparsity; MMV-CS; Structured sparse Bayesian learning | Common or partially overlapping sparse supports; Geometry-aligned delay-Doppler-angle structures | Effective with limited sensing resources; Exploits shared scattering structures for improved estimation | Support mismatch due to UE geometry and target visibility; Grid mismatch; Computational complexity |
| **Statistical Estimation** | Joint ML, MAP, Bayesian estimation and tracking | Shared target-state distributions, motion priors, environmental statistics, and inter-UE covariance structures. | Systematically incorporates prior knowledge and uncertainty | Statistical model mismatch; Weak or nonstationary correlation; Limited prior information; Computational complexity |
| **Subspace-Based Estimation** | Joint MUSIC, Distributed MUSIC, ESPRIT, Tensor-ESPRIT | Common subspaces, covariance structures, geometric constraints, Shared angular structures at BS | High-resolution estimation; mature technology, no sparse discretization | Requires sufficient SNR and snapshots; UE-dependent subspace mismatch; Limited ability to incorporate statistical priors and uncertainty models |
| **Learning-Assisted Estimation** | Deep unfolding, GNNs, physics-informed learning | Learned spatial, temporal, and geometric relationships, Shared environmental and target representations | Captures complex correlations and nonlinearities | Requires training data and generalization capability; Limited interpretability |

*Table 2 Comparison of Techniques that can exploit correlation between UEs for sensing.*

density, completeness, spatial resolution, and reconstruction metrics such as the Chamfer distance.

Multi-view fusion can operate at different abstraction levels [2, 3], with different sensing information and reporting requirements. Measurement-level fusion combines estimated sensing parameters or target-related measurements from multiple UEs, whereas feature-level fusion integrates representations such as range-Doppler maps, occupancy maps, or learned embeddings. Semantic-level fusion [4] further combines task-relevant information, such as object identities, activities, or scene descriptions. These approaches offer different trade-offs between sensing information preservation, fusion performance, and reporting overhead, particularly for downlink-assisted sensing. Specifically, higher-level abstraction may discard more detailed target signatures; their choice should therefore be task-dependent, balancing the preservation of relevant sensing information against communication overhead.

Multi-view fusion can be realized using a variety of techniques depending on the fusion level and the available prior information. Measurement- and feature-level fusion are often formulated as probabilistic inference problems, where observations collected from multiple UEs are jointly processed to estimate target states or environmental maps. Such fusion can be implemented through Bayesian inference, factor graphs, and message-passing frameworks that have been extensively studied in cooperative localization and distributed inference systems [12]. More recently, graph-based and learning-assisted fusion approaches have attracted growing interest due to their ability to capture complex relationships among heterogeneous sensing observations and network topologies. Developing scalable fusion frameworks that effectively integrate heterogeneous sensing information across multiple abstraction levels remains an important research direction for future multi-UE sensing systems.

## 4. Challenges and Open Research Problems

Multi-UE sensing remains at an early stage of development, and many fundamental challenges must be addressed before it can be practically deployed. Many challenges identified for cooperative ISAC networks [6] remain relevant but are further complicated by the large number and dynamic nature of participating UEs. These challenges are closely interconnected: synchronization must preserve inter-UE sensing correlations, which underpin joint parameter estimation and multi-view fusion, while their performance guides UE grouping and resource optimization. In turn, sensing results can inform subsequent estimation, fusion, and resource allocation. Table 3 provides an overview of the major open research challenges in multi-UE sensing, several of which are elaborated upon in the following subsections.

| Research Topics | Key Challenges | Potential Approaches | Impact on System |
|---|---|---|---|
| **Correlation Modelling** | Quantifying inter-UE sensing correlation under realistic geometries and mobility | Geometric modelling, stochastic geometry, ray tracing, learning-based modelling | Foundation for estimation, fusion, grouping, and resource allocation |
| **Performance Characterization** | Understanding achievable gains and multi-target separability | CRB, Ziv-Zakai bounds, information-theoretic analysis | Guides system design and UE deployment |
| **Joint Parameter Estimation, Association and Tracking** | Coupling between estimation and association; Missed detections and false alarms; Partial target visibility; Heterogeneous UE sensing quality | Signal-level correlation exploitation; Joint association and localization; Iterative association-estimation with target-state feedback; Multiple Hypothesis Tracking (MHT) and Random Finite Set (RFS) filters | Robust parameter estimation, target association, localization, and tracking under incomplete and heterogeneous observations |
| **Sensing Information Compression** | Reducing UE feedback overhead while preserving sensing utility | Compressive sensing, semantic sensing, distributed source coding, task-oriented compression | Scalability of downlink and hybrid sensing |
| **UE Grouping and Resource Optimization** | Selecting the most informative UEs and resources; Balancing sensing correlation and geometric diversity; Inter-cell interference and network-wide comm.-sensing trade-offs | Correlation-aware graph clustering and scheduling; Fisher-information-based selection; Interference-aware power control, beamforming, and coordinated resource allocation; Learning-assisted optimization | Improves sensing efficiency and network utilization; Enhanced target observability and separability; Mitigates inter-cell interference |
| **UE Localization** | Obtaining accurate UE positions, orientations, and synchronization states for sensing | Cooperative localization, anchor-assisted positioning, joint Tx-target localization | Enables target localization, imaging, association, fusion, and correlation exploitation |
| **Security and Privacy** | Preventing malicious sensing reports and protecting user information | Trust management, authentication, privacy-preserving sensing | Reliable and trustworthy sensing services |

*Table 3 Major Open Research Problems in Multi-UE Sensing.*

### 4.1 Correlation Modelling and Performance Characterization

Multi-UE sensing requires correlation models that characterize the relationships among observations from distributed UEs. These relationships depend on UE distribution, target states, sensing geometries, propagation conditions, and multiple-access resources. Representative models include deterministic geometric relationships, joint statistical distributions, and covariance structures, developed through analytical modelling, ray tracing, or measurements. Their effectiveness depends on target visibility, environmental dynamics, and model accuracy. Such models underpin correlation-aware estimation, fusion, and resource optimization.

Beyond modelling, a rigorous theoretical framework is needed to characterize the sensing performance achievable by multi-UE sensing systems. Important questions include how sensing accuracy and target resolution scale with the number and spatial distribution of UEs, the achievable coherent gains from bandwidth aggregation and distributed aperture formation under practical synchronization and calibration constraints, and the tradeoffs among sensing coverage, resolution, robustness, and communication overhead. Performance bounds, such as the Cramér-Rao lower bound and information-theoretic metrics, can provide valuable insights into the benefits and limitations of multi-view sensing. Establishing such theoretical foundations will not only guide system design but also reveal the fundamental gains achievable through correlation exploitation and multi-view fusion in future perceptive mobile networks.

### 4.2 Sensing Parameter Association and Tracking

A key challenge in multi-UE sensing is identifying common targets across UE observations and associating their estimated sensing parameters with physical targets, unless such associations are inherently resolved during joint estimation [10]. Two related problems arise: discriminating target-related paths from undesired environmental reflections, and associating multiple desired sensing paths with their corresponding physical targets, particularly for extended targets with multiple scattering points. Furthermore, because each UE forms a distinct bistatic sensing geometry and may only observe a subset of targets, sensing measurements collected across the network are often incomplete and partially overlapping. Importantly, the identification and exploitation of common targets can be performed centrally at the BS through correlation-aware estimation and target association, rather than at individual UEs.

To address the challenge of parameter association, clutter suppression and discrimination can be performed during signal processing by exploiting Doppler characteristics, background channel statistics, target signatures, and multi-view consistency. Correlation-aware joint estimation may further facilitate partial association of target-related paths during parameter estimation itself, as exemplified in [10]. Systems may also exploit geometric consistency among estimated sensing parameters, such as delay-Doppler-angle-pathloss relationships and common target localization constraints, to associate observations across UEs. For multiple desired sensing paths, spatial clustering, motion consistency, and temporal continuity may further facilitate their association with individual physical targets [2, 13].

Integrating such signal-level estimation with joint association and localization techniques [7] enables an iterative association-estimation framework. Initial sensing parameters and their uncertainties can guide probabilistic association and target-state estimation through geometric and motion consistency. The resulting association probabilities and target states can then provide priors for refining sensing parameters directly from multi-UE signals or CSI, followed by updated association and target-state estimation.

### 4.3 Sensing Information Compression at UEs

In downlink-assisted and hybrid multi-UE sensing, a potentially large number of UEs may participate in sensing, creating significant communication overhead when transmitting sensing information to the network for fusion. Consequently, efficient sensing information compression becomes essential for scalable multi-UE sensing.

Unlike conventional source compression, the objective is to preserve information relevant to subsequent sensing tasks rather than reconstruct the original measurements. Potential approaches include compressive sensing, low-dimensional feature extraction, task-oriented compression, and semantic sensing. Distributed source coding and correlation-aware feature compression can further reduce redundancy among UE reports, while joint optimization of compression and UE grouping can preserve complementary sensing information for fusion. Existing studies on CSI compression for large-scale WiFi sensing [14] provide useful insights, but compression schemes explicitly accounting for multi-view correlation and fusion performance remain largely unexplored.

### 4.4 UE Grouping, Communication-Sensing Resource Coupling and Optimization

A unique challenge in multi-UE sensing is UE grouping and participation selection. Since sensing contributions vary with UE locations, sensing geometries, propagation conditions, and target visibility, activating all available UEs is often neither necessary nor efficient. A distinctive consideration is the interplay between inter-UE correlation and geometric diversity. Geographically proximate UEs may provide strongly correlated observations that facilitate joint parameter estimation, whereas more widely distributed UEs may offer complementary geometries that enhance target observability and separability. For example, in RF point cloud reconstruction, the former may improve scattering point localization and separability, while the latter may reveal additional scattering points. Consequently, point cloud density may exhibit diminishing returns as the observations of additional UEs increasingly overlap. UE grouping should therefore jointly consider sensing correlation, geometric diversity, target observability, and predicted sensing contributions to maximize sensing utility while limiting communication overhead and computational complexity.

These considerations motivate joint communication-sensing optimization through correlation-aware clustering, information-driven UE selection, and learning-assisted optimization of UE grouping, sensing resources, and information reporting. Such optimization relies on correlation modelling and performance characterization to quantify individual UE contributions and predict collective sensing gains. Furthermore, sensing requirements may be incorporated into communication resource management [5] to adapt sensing coverage, tracking accuracy, and environmental awareness to network conditions and application demands. Developing scalable cross-layer optimization frameworks integrating correlation modelling, performance characterization, UE grouping, resource allocation, and sensing fusion remains an important research direction.

### 4.5 Other Open Research Problems

Other challenges include UE localization, security, privacy, and protocol support. Accurate UE positions, orientations, and synchronization states are often required for absolute sensing parameter recovery, motivating cooperative positioning, anchor-assisted calibration, and joint UE-target localization [**7**, **12**]. Distributed sensing also introduces vulnerabilities to spoofed reports, adversarial attacks, and leakage of location or environmental information [**15**]. Practical deployment further requires support for heterogeneous UE capabilities, sensing-information reporting, uncertainty metadata, privacy permissions, and resource scheduling. Addressing these issues is essential for reliable large-scale multi-UE sensing.

## 5. Conclusion

This article presented a systematic framework for multi-UE networked sensing in future perceptive mobile networks. A central insight is that distributed UE observations offer both exploitable sensing correlations and complementary geometries, enabling joint processing to overcome the limitations of individual sensing links. Moreover, collectively exploiting time, frequency, and spatial communication resources creates new opportunities for sensing performance enhancement. Realizing these gains requires correlation-preserving synchronization, scalable joint estimation and fusion, and sensing-aware UE grouping and resource optimization. Addressing their interdependencies will be crucial to enabling robust, scalable, and network-native environmental perception in future ISAC systems.